\documentclass[a4paper]{article}

\usepackage{fullpage}

\usepackage{color}
\usepackage{graphicx}
\usepackage{amsmath}
\usepackage{amssymb}
\usepackage{amsthm}
\usepackage{tikz}
\usepackage{enumitem}
\usepackage{booktabs}
\usepackage{tabularx}
\usepackage{array}
\usepackage[hidelinks]{hyperref}
\usepackage{cleveref}

\usepackage{algorithm}
\usepackage{algpseudocode}

\algrenewcommand\algorithmicrequire{\textbf{Input:}}
\algrenewcommand\algorithmicensure{\textbf{Output:}}

\theoremstyle{plain}
\newtheorem{theorem}{Theorem}
\newtheorem{lemma}[theorem]{Lemma}

\newtheorem{corollary}[theorem]{Corollary}

\theoremstyle{definition}

\theoremstyle{remark}
\newtheorem{remark}[theorem]{Remark}

\def\eps{\varepsilon}
\newcommand{\IR}{\mathbb{R}}
\newcommand{\HH}{\mathbb{H}}
\newcommand{\SSph}{\mathbb{S}}

\newcommand{\diam}{\operatorname{diam}}
\newcommand{\rad}{\operatorname{rad}}
\newcommand{\conv}{\operatorname{conv}}
\newcommand{\arcosh}{\operatorname{arcosh}}

\newcommand{\Hd}{d_{\mathrm H}}

\title{Optimal Coresets for Hyperbolic Farthest-Point Queries via Ideal-Boundary Envelopes
\thanks{
This work was supported by National Research Foundation of Korea (NRF) grants funded by the Korea government (MSIT) (No. RS-2026-25471649 and No. RS-2024-00414849).
}}

\author{Eunku Park\\
Department of Liberal Arts and Sciences\\
DGIST, Republic of Korea\\
\texttt{parkeun9@dgist.ac.kr}
}

\begin{document}
\maketitle

\begin{abstract}
We study coresets for farthest-point queries in hyperbolic space. Given a nonempty finite set $P \subset \HH^D$ and $0<\eps \le 1$, we seek a coreset $P_{\eps} \subseteq P$ whose farthest distance from every query point underestimates that of $P$ by at most an additive $\eps$ and retains at least a $1-\eps$ fraction of it. For every fixed $D \ge 2$, we prove that the optimal worst-case coreset size is $\Theta\bigl(\eps^{-(D-1)/2}\bigr)$.

Our main geometric ingredient is an exact reduction from queries throughout hyperbolic space to an upper envelope on its ideal boundary. In the hyperboloid model, each input point induces a positive boundary-score function whose logarithm records its asymptotic distance offset along geodesic rays from the origin. We call the pointwise maximum of these functions the \emph{ideal-boundary envelope} of the input set. Our structure theorem states that the supremum of the additive farthest-distance loss over all query points is exactly the maximum logarithmic gap between the input and coreset envelopes.

For the upper bound, we move the minimum-enclosing-ball center to the origin and scale the spatial coordinates. This normalization produces a bounded Euclidean point set and keeps its boundary envelope bounded away from zero. For each ideal direction, the normalized score is a convex, uniformly Lipschitz function of the spatial coordinates in $\IR^D$. A standard Euclidean kernel therefore approximates all directional score maxima simultaneously. The structure theorem transfers this approximation to both required guarantees. For the lower bound, we place input points on a fixed-radius hyperbolic sphere using a spherical packing and choose corresponding antipodal queries. The hyperbolic cosine law makes every input point indispensable, establishing the lower bound even when either guarantee is required separately.
\end{abstract}


\section{Introduction}\label{sec:introduction}
Given a finite point set $P$ in a metric space, a farthest-point query asks for a point of $P$ farthest from a query point $x$. For a nonempty finite set $A \subset \HH^D$, let 
\[
 F_A(x)=\max_{a\in A} d_{\mathrm H}(x,a)
\]
denote its farthest-distance function. The goal of a coreset is to approximate $F_P$ uniformly over all query points while retaining only a small number of input points.

In fixed-dimensional Euclidean space, approximate farthest-point queries are closely related to directional extrema and Euclidean $\eps$-kernels~\cite{AgarwalMatousekSuri1992, AgarwalHarPeledVaradarajan2004}. Euclidean approximation guarantees are naturally stated relative to the diameter or directional width of the input because a dilation multiplies every distance and every absolute approximation loss by the same factor. Hyperbolic space admits no analogous global similarity that uniformly rescales all distances. Once its standard metric is fixed, an absolute additive error therefore has an intrinsic geometric scale.

Park and Vigneron~\cite{ParkVigneron2025} initiated the study of coresets for farthest-point queries in fixed-dimensional hyperbolic space. They constructed a coreset of size $O(\eps^{-D})$ that, for every query point, simultaneously provides additive error at most $\eps$ and multiplicative factor at least $1-\eps$. Their proof treats bounded and unbounded geometric scales separately, using local grids, thin triangles, and angular decompositions. This leaves the natural question of determining the optimal dependence on $\eps$.

We resolve this question by proving that the optimal exponent is $(D-1)/2$. This is the same worst-case exponent as for Euclidean $\eps$-kernels and Euclidean farthest-point coresets under their standard scale-invariant guarantees~\cite{AgarwalMatousekSuri1992, AgarwalHarPeledVaradarajan2004}. The exponent reflects the fact that the relevant directions form the $(D-1)$-dimensional ideal boundary and that a directional deviation of angle $\theta$ produces a second-order loss $1-\cos\theta=\Theta(\theta^2)$. Thus, approximation error $\eps$ corresponds to angular resolution on the order of $\sqrt{\eps}$, and covering the ideal boundary at this resolution requires $\Theta \left((\sqrt{\eps})^{-(D-1)}\right) = \Theta \left(\eps^{-(D-1)/2}\right)$ directions. Although the Euclidean and hyperbolic bounds have the same dependence on $\eps$, their additive guarantees use different scales. The Euclidean guarantee is naturally normalized by the diameter or directional width of the input, whereas our hyperbolic coreset provides an absolute additive guarantee in the fixed hyperbolic metric. Our proof makes this intuition precise through an exact representation on the ideal boundary and a Euclidean convex-hull approximation.

\paragraph{Main result.}
A nonempty set $P_{\eps}\subseteq P$ is an \emph{additive $\eps$-coreset} if $F_{P_{\eps}}(x) \ge F_P(x)-\eps$ for every $x \in \HH^D$, and it is a \emph{multiplicative $\eps$-coreset} if $F_{P_{\eps}}(x) \ge (1-\eps)F_P(x)$ for every $x \in \HH^D$. We call $P_{\eps}$ an $\eps$-coreset when it satisfies both guarantees.

\begin{theorem}\label{thm:main}
Fix an integer $D \ge 2$. There are positive constants $c_D$ and $C_D$ such that the following statements hold.
\begin{enumerate}
 \item For every nonempty finite set $P \subset \HH^D$ and every  $0 < \eps \le 1$, there is an $\eps$-coreset $P_{\eps}\subseteq P$ satisfying, simultaneously for all $x \in \HH^D$,
 \begin{equation}\label{eq:simultaneous-guarantees}
  F_{P_{\eps}}(x) \ge F_P(x)-\eps \qquad\text{and}\qquad F_{P_{\eps}}(x) \ge (1-\eps)F_P(x),
 \end{equation}
 and
 \[
  |P_{\eps}| \le C_D\eps^{-(D-1)/2}.
 \]

 \item For every $0 < \eps \le 1/36$, there is a finite set  $P^{\star}\subset\HH^D$ with
 \[
  |P^{\star}| \ge c_D\eps^{-(D-1)/2}
 \]
 such that no proper subset of $P^{\star}$ is an additive $\eps$-coreset of $P^{\star}$, and no proper subset is a multiplicative $\eps$-coreset of $P^{\star}$.
\end{enumerate}
Consequently, for every fixed $D \ge 2$, the worst-case minimum size of a coreset satisfying both requirements is $\Theta \left(\eps^{-(D-1)/2}\right)$.
\end{theorem}

For $D=1$, the hyperbolic line is isometric to the real line. Hence, the two extreme input points form an exact coreset, since every query point is farthest from one of these two points. If $P$ consists of a single point, that point alone is an exact coreset. Thus, the nontrivial dependence on $\eps$ begins in dimension $D=2$.

The upper bound is constructive. Combining our hyperbolic reduction with the optimal-size Euclidean $\eps$-kernel framework of Agarwal, Har-Peled, and Varadarajan~\cite{AgarwalHarPeledVaradarajan2004} and the efficient
construction algorithm of Arya, da Fonseca, and Mount~\cite{AryaDaFonsecaMount2017} gives the following algorithmic
consequence. We state the running time separately because Theorem~\ref{thm:main} concerns the optimal worst-case coreset size, whereas the algorithmic bound additionally depends on the chosen kernel construction and the computational model.

\begin{theorem}\label{thm:algorithmic}
For every fixed $D \ge 3$ and every fixed constant $\alpha>0$, let $P \subset \HH^D$ be a set of $n$ points. In the real-RAM model, a coreset satisfying Theorem~\ref{thm:main}(1) can be constructed in expected time
\[
 O\left( n\log\frac{2}{\eps}+\eps^{-(D-1)/2-\alpha} \right).
\]
Here, $\alpha$ is an arbitrarily small exponent slack inherited from the Euclidean $\eps$-kernel construction of Arya, da Fonseca, and Mount~\cite{AryaDaFonsecaMount2017}.
\end{theorem}

The Euclidean kernel construction used in Theorem~\ref{thm:algorithmic} is deterministic. The expectation arises only from computing the hyperbolic minimum enclosing ball. By Lemma~\ref{lem:meb-lp-type}, this is an LP-type problem of combinatorial dimension at most $D+1$. Hence, for fixed $D$, it can be computed in expected linear time by the standard randomized incremental algorithm~\cite{Welzl1991,SharirWelzl1992}.

\paragraph{Our approach.}
Our proof has three main steps. First, we replace the approximation problem over all query points in hyperbolic space with an approximation problem on its ideal boundary. In the hyperboloid model, every input point induces a positive score in each ideal direction. The logarithm of this score records the constant distance offset observed by a query moving to infinity in that direction. Taking the pointwise maximum of these scores over a finite set produces what we call its \emph{ideal-boundary envelope}. We prove that the largest farthest-distance loss over all finite query points is exactly the largest logarithmic loss between the corresponding ideal-boundary envelopes. The proof extends the geodesic segment from a farthest point through the query point. As the query moves along the resulting ray, its coreset loss cannot decrease, and at the ideal endpoint it converges to the envelope loss. Conversely, every envelope loss is approached by finite queries moving sufficiently far along the corresponding ray. This structure theorem allows us to replace the unbounded family of query points in $\HH^D$ with the $(D-1)$-dimensional family of ideal directions.

For the upper bound, we normalize the input using its hyperbolic minimum enclosing ball. We move the center of this ball to the hyperboloid origin and scale the spatial coordinates by the exponential of the negative enclosing radius. The elementary optimality condition for the enclosing ball shows that the directions of the input points lying on its boundary balance around the origin. Together with the scaling, this implies that the normalized point set has controlled Euclidean diameter and that its ideal-boundary envelope remains bounded away from zero. More importantly, every normalized boundary score becomes the value of a convex, $2$-Lipschitz function on a point of $\IR^D$. A standard Euclidean kernel with approximation parameter proportional to $\eps$ therefore approximates all of these scores simultaneously. The structure theorem transfers this approximation back to every hyperbolic query and bounds the farthest-distance loss by $\eps\min\{R,1\}$, where $R$ is the minimum-enclosing-ball radius. This is at most $\eps$, and it is also at most $\eps F_P(x)$ for every query $x$, because $F_P(x) \ge R$. Hence the same coreset satisfies both the additive and multiplicative guarantees. The Euclidean kernel theorem gives a coreset of size $O\bigl(\eps^{-(D-1)/2}\bigr)$.

For the lower bound, we choose directions on $\SSph^{D-1}$ separated by an angle proportional to $\sqrt{\eps}$ and place one input point at hyperbolic radius one in each direction. For every input point, the point at radius one in the antipodal direction serves as a query at which that input point is uniquely farthest. The hyperbolic cosine law shows that every other input point is closer by more than the error allowed by either approximation guarantee. Consequently, every input point is indispensable. A spherical packing at angular scale $\sqrt{\eps}$ contains $\Omega\bigl(\eps^{-(D-1)/2}\bigr)$ directions, which proves the lower bound and establishes the optimal coreset size.

\paragraph{Relation to previous work.}
The most directly related previous result is the $O(\eps^{-D})$-size hyperbolic coreset of Park and Vigneron~\cite{ParkVigneron2025}. A direct application of a Euclidean kernel theorem to the hyperboloid coordinates would not yield our sharper bound. Such an approach would treat the input as an arbitrary point set in $\IR^{D+1}$. Since a standard $\eps$-kernel in dimension $d$ has size $O(\eps^{-(d-1)/2})$, it would give only $O(\eps^{-D/2})$ points.

Our improvement relies on two additional ingredients. First, the ideal-boundary envelope identity converts the worst coreset loss over all hyperbolic query points exactly into a uniform approximation problem on the ideal boundary. Second, after normalizing with respect to the minimum enclosing ball, each boundary score becomes a convex, Lipschitz function of the $D$-dimensional spatial coordinate. This allows us to apply the Euclidean kernel in $\IR^D$ rather than in $\IR^{D+1}$, giving $O(\eps^{-(D-1)/2})$ points. The lower bound proves that this dependence on $\eps$ is optimal.

For Euclidean farthest-point queries, Agarwal, Matou\v{s}ek, and Suri~\cite{AgarwalMatousekSuri1992} used directional extrema to obtain approximate farthest neighbors. The theory of $\eps$-kernels and its applications to geometric extent measures was developed by Agarwal, Har-Peled, and Varadarajan~\cite{AgarwalHarPeledVaradarajan2004}; see also Har-Peled's monograph~\cite{HarPeled2011}. The size bound $O\left(\eps^{-(D-1)/2}\right)$ is rooted in classical convex-body approximation results of Dudley~\cite{Dudley1974} and Bronshteyn and Ivanov~\cite{BronshteynIvanov1975}. Efficient algorithms for constructing kernels of optimal size were given by Arya, da Fonseca, and Mount~\cite{AryaDaFonsecaMount2017}.

The logarithm of the ideal-boundary envelope is closely related to horospherical support functions in hyperbolic convex geometry. Such functions appear in the hyperbolic Minkowski theory of Li, Wan, and Xu~\cite{LiWanXu2024}. Fletcher et al.~\cite{FletcherMoellerPhillipsVenkatasubramanian2011} study horoball hulls and approximations of geometric extent. Our contribution is an exact identity relating the uniform farthest-distance loss over all query points to the logarithmic loss between ideal-boundary envelopes, together with optimal bounds on coreset size. We use the term \emph{ideal-boundary envelope} because, in each ideal direction, the envelope takes the maximum of the boundary scores contributed by the input points.

Standard background on the hyperboloid model and its ideal boundary may be found in Ratcliffe~\cite{Ratcliffe2006} and Bridson and Haefliger~\cite{BridsonHaefliger1999}. All hyperbolic estimates needed for our main proof are derived directly from the hyperboloid distance formula.

The remainder of the paper is organized as follows. Section~\ref{sec:preliminaries} reviews the required hyperbolic and Euclidean tools. Section~\ref{sec:structure} proves the structure theorem. Sections~\ref{sec:upper-bound} and~\ref{sec:algorithm} establish the upper bound and its algorithmic implementation. Section~\ref{sec:lower-bound} proves the lower bound, and Section~\ref{sec:conclusion} concludes the paper.


\section{Preliminaries}\label{sec:preliminaries}
\subsection{The hyperboloid model}
Throughout this paper, $\langle\cdot,\cdot\rangle$ and $\|\cdot\|$ denote the Euclidean inner product and norm on $\IR^D$, respectively, and $\SSph^{D-1}=\{u\in\IR^D:\|u\|=1\}$ denotes the Euclidean unit sphere. We use the standard normalization of hyperbolic space with sectional curvature $-1$.

The hyperboloid model represents a hyperbolic point by a vector $x=(x_0,\bar x)\in\IR^{D+1}$, where $x_0\in\IR$ is the first coordinate and $\bar x\in\IR^D$ is the spatial coordinate. For two such vectors, define the \emph{Lorentz bilinear form} by $\langle x,y \rangle_{\mathrm L} = x_0y_0-\langle\bar x,\bar y\rangle$.
The model consists of the upper sheet
\[
 \HH^D =\left\{x=(x_0,\bar x)\in\IR^{D+1} \mathrel{\big|}x_0^2-\|\bar x\|^2=1,\ x_0>0\right\}.
\]
Every point on this sheet satisfies $x_0=\sqrt{1+\|\bar x\|^2}$. Thus, the spatial coordinate determines the first coordinate uniquely, and the model is $D$-dimensional despite its representation in $\IR^{D+1}$. For $x,y\in\HH^D$, the hyperbolic distance $\Hd(x,y)$ is determined by
\begin{equation}\label{eq:hyperboloid-distance}
 \cosh\Hd(x,y) = \langle x,y\rangle_{\mathrm L} = x_0y_0-\langle\bar x,\bar y\rangle.
\end{equation}
All distances in $\HH^D$ are measured using this metric. The ambient Euclidean distance between the representing vectors is a different quantity. The model and its metric are described in~\cite[Section~3.2]{Ratcliffe2006}.

We write $o=(1,\mathbf 0)$ for the distinguished origin, where $\mathbf 0$ is the zero vector in $\IR^D$. An \emph{isometry} of $\HH^D$ is a bijection that preserves hyperbolic distances. To describe the isometries used below, a \emph{Lorentz transformation} is an invertible linear map $\Phi:\IR^{D+1}\to\IR^{D+1}$ satisfying $\langle\Phi x,\Phi y\rangle_{\mathrm L} = \langle x,y\rangle_{\mathrm L}$ for all $x,y \in \IR^{D+1}$; see~\cite[Section~3.1]{Ratcliffe2006}. If $\Phi$ maps the upper sheet onto itself, Equation~\eqref{eq:hyperboloid-distance} shows that its restriction to $\HH^D$ is an isometry \cite[Theorem~3.2.3]{Ratcliffe2006}. For any $c\in\HH^D$, one can choose such a transformation with $\Phi(c)=o$ \cite[Theorems~3.1.6 and~3.2.3]{Ratcliffe2006}. We may therefore move any chosen point to the origin while preserving all distances.

For $p=(p_0,\bar p)\in\HH^D$, put $r=\Hd(o,p)$. The distance formula gives $p_0=\cosh r$, and the hyperboloid equation then gives $\|\bar p\|=\sinh r$. Consequently, every $p\ne o$ has unique polar coordinates
\begin{equation}\label{eq:polar-coordinate} 
 p=(\cosh r,\sinh r\,u), \qquad r=\Hd(o,p)>0, \qquad u=\frac{\bar p}{\|\bar p\|}\in\SSph^{D-1}.
\end{equation}
Here $r$ is the hyperbolic distance from the origin, and $u$ specifies the spatial direction. At $p=o$, we have $r=0$, and any choice of $u$ represents the same point.

A \emph{unit-speed geodesic ray} is a map $\gamma:[0,\infty)\to\HH^D$ satisfying $\Hd(\gamma(s),\gamma(t))=|s-t|$ for all $s,t \ge 0$. Thus, its parameter measures hyperbolic distance traveled from its starting point. To obtain a ray from $o$ in a direction $\xi\in\SSph^{D-1}$, consider points of the form $(a,b\xi)$. The hyperboloid equation becomes $a^2-b^2=1$, which leads to the parametrization
\begin{equation}\label{eq:ideal-ray}
 \gamma_\xi(t)=(\cosh t,\sinh t\,\xi), \qquad t\ge0.
\end{equation}
This curve starts at $o$ and lies on the hyperboloid. Moreover,
\[
 \cosh\Hd(\gamma_\xi(s),\gamma_\xi(t)) = \cosh s\cosh t-\sinh s\sinh t = \cosh(t-s).
\]
Since distance is nonnegative, this identity gives $\Hd(\gamma_\xi(s),\gamma_\xi(t))=|t-s|$. Thus $\gamma_\xi$ is a unit-speed geodesic ray. Every unit-speed geodesic ray from $o$ has this form~\cite[Theorem~3.2.5]{Ratcliffe2006}. Allowing $t$ to range over all of $\IR$ gives a complete geodesic line. More generally, any two distinct points in $\HH^D$ are joined by a unique geodesic segment, which lies on a unique complete geodesic line \cite[Theorems~3.2.4--3.2.5]{Ratcliffe2006}. We will use this extension property in the structure theorem.

The \emph{ideal boundary}, denoted by $\partial_\infty\HH^D$, records the possible directions of escape to infinity. Formally, two unit-speed geodesic rays $\rho$ and $\sigma$ are called \emph{asymptotic} if $\sup_{t\ge0}\Hd(\rho(t),\sigma(t))<\infty$. The ideal boundary consists of their equivalence classes under this relation~\cite[Definition~II.8.1]{BridsonHaefliger1999}. Each class has exactly one representative ray starting at $o$
\cite[Proposition~II.8.2]{BridsonHaefliger1999}. These representatives are precisely the rays $\gamma_\xi$, so they give a bijection $\SSph^{D-1}\longrightarrow\partial_\infty\HH^D$ using $\xi\longmapsto[\gamma_\xi]$, where $[\gamma_\xi]$ denotes the equivalence class containing $\gamma_\xi$. We use this bijection to identify the ideal boundary with $\SSph^{D-1}$. In particular, $\xi$ specifies an ideal endpoint, rather than a point of $\HH^D$. Rays with different starting points may have the same ideal endpoint; the representative starting at $o$ is unique.

For a nonempty finite set $A\subset\HH^D$, define its \emph{farthest-distance function} by $F_A(x)=\max_{a\in A}\Hd(x,a)$. A closed hyperbolic ball centered at $x$ contains $A$ exactly when its radius is at least $F_A(x)$. The \emph{minimum-enclosing-ball radius} is therefore $\rad(A)=\min_{x\in\HH^D}F_A(x)$. This minimum is attained at a unique point, by the circumcenter theorem applied to hyperbolic space \cite[Proposition~II.2.7]{BridsonHaefliger1999}. We call this point the \emph{minimum-enclosing-ball center}, or the \emph{circumcenter}, of $A$. The elementary optimality condition needed for our construction is proved in the upper-bound section.

\subsection{Euclidean \texorpdfstring{$\delta$}{delta}-kernels}
The standard notion of a Euclidean $\eps$-kernel preserves the directional widths of a point set up to a prescribed relative error~\cite{AgarwalHarPeledVaradarajan2004}. We denote this error parameter by $\delta$ throughout the paper and use the term \emph{$\delta$-kernel}, reserving $\eps$ for the accuracy of the hyperbolic coreset.

Let $Z\subset\IR^D$ be a nonempty finite set. We write $\conv Z$ for its convex hull and $\diam(Z)=\max_{z,w\in Z}\|z-w\|$ for its Euclidean diameter. For a direction $v\in\SSph^{D-1}$, define the \emph{support value} $H_Z(v)=\max_{z\in Z}\langle v,z\rangle$ and the \emph{directional width}
\[
 w_Z(v) = \max_{z\in Z}\langle v,z\rangle -\min_{z\in Z}\langle v,z\rangle = H_Z(v)+H_Z(-v).
\]
Geometrically, $w_Z(v)$ is the distance between the two supporting hyperplanes of $\conv Z$ perpendicular to $v$. For $0<\delta<1$, a nonempty subset $Z'\subseteq Z$ is a \emph{$\delta$-kernel} of $Z$ if $w_{Z'}(v)\ge(1-\delta)w_Z(v)$ for every $v \in \SSph^{D-1}$. Since $Z'\subseteq Z$, we also have $w_{Z'}(v)\le w_Z(v)$.

Our upper-bound proof uses the resulting approximation of the convex hull. For nonempty compact sets $K'\subseteq K\subset\IR^D$, their Euclidean Hausdorff distance reduces to
\[
 d_{\mathrm H}^{\mathrm E}(K,K') = \max_{x\in K}\min_{y\in K'}\|x-y\|.
\]
Thus, this distance is at most $\eta$ exactly when every point of $K$ is within Euclidean distance $\eta$ of $K'$.

The following standard kernel theorem provides the size and construction-time bounds that we need. The size bound has its origins in classical convex-body approximation \cite{Dudley1974,BronshteynIvanov1975}; the formulation for finite point sets appears in the theory of extent measures \cite{AgarwalHarPeledVaradarajan2004}. The construction-time bound is due to Arya, da Fonseca, and Mount~\cite[Theorem~1.1]{AryaDaFonsecaMount2017}. We derive the stated Hausdorff bound below.

\begin{theorem}[Euclidean $\delta$-kernel theorem]\label{thm:euclidean-kernel}
For every fixed $D \ge 2$, every nonempty finite set $Z \subset \IR^D$, and every $0<\delta<1$, there exists a $\delta$-kernel $Z'\subseteq Z$ of size $|Z'|=O\left(\delta^{-(D-1)/2}\right)$. Moreover,
\begin{equation}\label{eq:hausdorff-kernel}
 d_{\mathrm H}^{\mathrm E}(\conv Z,\conv Z') \le\delta\,\diam(Z).
\end{equation}
For every fixed $D \ge 3$ and every fixed constant $\alpha>0$, such a kernel can be constructed deterministically in the
real-RAM model in time
\[
 O\left( |Z|\log\frac{2}{\delta} +\delta^{-(D-1)/2-\alpha} \right).
\]
\end{theorem}

Here $\alpha$ is an arbitrarily small fixed exponent slack in the construction time. It does not affect the kernel size.

For completeness, we prove~\eqref{eq:hausdorff-kernel} directly from the directional-width guarantee. Define $\Delta(v)=H_Z(v)-H_{Z'}(v)$. Both $\Delta(v)$ and $\Delta(-v)$ are nonnegative because $Z'\subseteq Z$. Therefore,
\[
 0 \le \Delta(v) \le \Delta(v)+\Delta(-v) = w_Z(v)-w_{Z'}(v) \le \delta w_Z(v) \le \delta\,\diam(Z).
\]

Put $K=\conv Z$ and $K'=\conv Z'$. Fix $x\in K\setminus K'$, and let $y$ be a closest point to $x$ in $K'$; it exists by compactness. Set $v=(x-y)/\|x-y\|$. For every $z\in K'$, convexity and the choice of $y$ imply $\langle v,z-y\rangle\le0$. Otherwise, moving a sufficiently small distance from $y$ toward $z$ would decrease the distance to $x$. Thus, $H_{Z'}(v)=\langle v,y\rangle$. Since $x\in\conv Z$, we also have $\langle v,x\rangle\le H_Z(v)$, and hence
\[
 \|x-y\| = \langle v,x-y\rangle \le H_Z(v)-H_{Z'}(v) \le\delta \diam(Z).
\]
Points of $K'$ have distance zero from $K'$. Since $K'\subseteq K$, this proves~\eqref{eq:hausdorff-kernel}.


\section{The Structure Theorem}\label{sec:structure}
This section relates farthest-point queries in $\HH^D$ to functions on its ideal boundary. Each input point defines a positive \emph{boundary score}, and the pointwise maximum of these scores is the \emph{ideal-boundary envelope} of the point set. We prove that the largest additive loss over all query points equals the largest logarithmic gap between the corresponding envelopes.

\subsection{Boundary scores and ideal-boundary envelopes}
For $p=(p_0,\bar p) \in \HH^D$ and $\xi \in \SSph^{D-1}$, define
\begin{equation}\label{eq:ell-score}
 \ell_p(\xi) = p_0-\langle\bar p,\xi\rangle = \langle p,(1,\xi)\rangle_{\mathrm L}.
\end{equation}
We call $\ell_p(\xi)$ the \emph{boundary score} of $p$ in direction $\xi$. It is positive because $\langle\bar p,\xi\rangle \le\|\bar p\| <\sqrt{1+\|\bar p\|^2} = p_0$.

The reason for this definition comes directly from the distance formula. Along the standard ray $\gamma_\xi(t)=(\cosh t,\sinh t\,\xi)$, we have
\begin{equation}\label{eq:standard-ray-score}
 \cosh\Hd(\gamma_\xi(t),p) = p_0\cosh t-\langle\bar p,\xi\rangle\sinh t = \frac{e^t}{2}\ell_p(\xi) + \frac{e^{-t}}{2}\ell_p(-\xi). 
\end{equation}
Thus $\ell_p(\xi)$ is the coefficient of $e^t/2$ in this expression for the hyperbolic cosine of the distance. Lemma~\ref{lem:ray-asymptotic} will show that $\log\ell_p(\xi)$ is the additive offset of the distance itself as the query moves far out along this ray.

For example, if $p=(\cosh r,\sinh r\,u)$ with $r>0$, then
\[
 \ell_p(\xi)=\cosh r-\sinh r\,\langle u,\xi\rangle, \qquad \ell_p(u)=e^{-r}, \qquad \ell_p(-u)=e^r.
\]
These two extreme scores reflect the distances along the geodesic through $o$ and $p$. A query beyond $p$ in direction
$u$ has distance $t-r$ from $p$, whereas a query in direction $-u$ has distance $t+r$.

For a nonempty finite set $A\subset\HH^D$, define its \emph{ideal-boundary envelope} by
\begin{equation}\label{eq:ideal-boundary-envelope}
 L_A(\xi)=\max_{a\in A}\ell_a(\xi), \qquad \xi\in\SSph^{D-1}.
\end{equation}
The function $L_A$ is the pointwise upper envelope of the boundary-score functions induced by the points of $A$. It is positive and continuous because it is the maximum of finitely many positive continuous functions.

\subsection{Distance estimates along geodesic rays}
We now give an explicit distance estimate for a ray with an arbitrary starting point. This is the form needed when we extend a geodesic through a given query point.

\begin{lemma}[Distance along a geodesic ray]\label{lem:ray-asymptotic}
Let $\rho$ be the unit-speed geodesic ray starting at $x \in \HH^D$ with ideal endpoint $\xi \in \SSph^{D-1}$. For every $a \in \HH^D$ and every $t \ge 0$,
\begin{equation}\label{eq:ray-asymptotic}
 0 \le \Hd(\rho(t),a)-t -\log\frac{\ell_a(\xi)}{\ell_x(\xi)} \le e^{2\Hd(x,a)-2t}.
\end{equation}
Consequently, for every nonempty finite set $A \subset \HH^D$,
\begin{equation}\label{eq:farthest-ray-asymptotic}
 0 \le F_A(\rho(t))-t -\log\frac{L_A(\xi)}{\ell_x(\xi)} \le e^{2F_A(x)-2t}.
\end{equation}
In particular,
\[
 F_A(\rho(t))-t \longrightarrow \log L_A(\xi)-\log\ell_x(\xi) \qquad(t\to\infty).
\]
For the standard ray $\gamma_\xi$, the starting point is $o$ and $\ell_o(\xi)=1$.
\end{lemma}

\begin{proof}
We first write the ray explicitly using its starting point and ideal endpoint. Put $h=\ell_x(\xi)>0$ and define
\begin{equation}\label{eq:ray-from-query}
 r(t)=e^{-t}x+\frac{\sinh t}{h}(1,\xi), \qquad t\in\IR.
\end{equation}
The identities
\[
 \langle x,x\rangle_{\mathrm L}=1, \qquad \langle x,(1,\xi)\rangle_{\mathrm L}=h, \qquad \langle(1,\xi),(1,\xi)\rangle_{\mathrm L}=0
\]
give
\[
 \langle r(s),r(t)\rangle_{\mathrm L} = e^{-s-t}+e^{-s}\sinh t+e^{-t}\sinh s = \cosh(t-s).
\]
In particular, $\langle r(t),r(t)\rangle_{\mathrm L}=1$. The first coordinate of $r(t)$ cannot vanish, and it is positive at $t=0$; by continuity it is positive for every $t \in \IR$. Thus $r$ is a complete unit-speed geodesic with $r(0)=x$.

For $t\ge0$, another application of the distance formula yields
\[
 \cosh\Hd(r(t),\gamma_\xi(t)) = x_0 e^{-2t} + \frac{h+h^{-1}}{2}(1-e^{-2t}).
\]
This expression is bounded as $t$ varies over $[0,\infty)$. Hence $r$ and $\gamma_\xi$ are asymptotic and have the same ideal endpoint. The ray from $x$ to a given ideal endpoint is unique~\cite[Proposition~II.8.2]{BridsonHaefliger1999}, so $\rho(t) = r(t)$ for $t \ge 0$.

Fix $a\in\HH^D$, write $d_0=\Hd(x,a)$, and put $\alpha={\ell_a(\xi)}/{h}$, $\beta=2\cosh d_0-\alpha$. Equation~\eqref{eq:ray-from-query} gives, for every real $t$,
\begin{equation}\label{eq:general-ray-distance}
 \cosh\Hd(r(t),a) = \frac{\alpha e^t+\beta e^{-t}}{2}.
\end{equation}
We have $\alpha>0$. The right-hand side is at least one for every real $t$, so $\beta>0$ as well. Its minimum over
$t \in \IR$ is $\sqrt{\alpha\beta}$; therefore $\alpha\beta \ge 1$. Together with $\alpha+\beta=2\cosh d_0$, this implies $e^{-d_0}\le\alpha,\beta\le e^{d_0}$ and ${\beta}/{\alpha}\le e^{2d_0}$. Indeed, $\alpha\beta \ge 1$ gives $(\alpha-\cosh d_0)^2 \le \sinh^2d_0$, and the same argument applies to $\beta$.

We next estimate the distance in~\eqref{eq:general-ray-distance}.
Since $\alpha\beta \ge 1$,
\[
 \cosh\Hd(\rho(t),a) \ge \frac{\alpha e^t+\alpha^{-1}e^{-t}}{2} = \cosh(t+\log\alpha).
\]
It follows that $\Hd(\rho(t),a)\ge|t+\log\alpha|\ge t+\log\alpha$. For the upper estimate, use
\[
 \arcosh s=\log\bigl(s+\sqrt{s^2-1}\bigr)\le\log(2s), \qquad s\ge1.
\]
Combining this inequality with $\log(1+u)\le u$ for $u\ge0$ gives
\[
 \Hd(\rho(t),a) \le \log\bigl(\alpha e^t+\beta e^{-t}\bigr) = t+\log\alpha +\log\left(1+\frac{\beta}{\alpha}e^{-2t}\right) \le t+\log\alpha+e^{2d_0-2t}.
\]
This proves~\eqref{eq:ray-asymptotic}.

Finally, $\Hd(x,a)\le F_A(x)$ for every $a \in A$. Taking the maximum over $a\in A$ in the two distance bounds, and using $\max_{a\in A}\log\ell_a(\xi)=\log L_A(\xi)$, proves~\eqref{eq:farthest-ray-asymptotic}.
\end{proof}

The term $-\log\ell_x(\xi)$ depends on the starting point and ideal direction, but not on the input point or point set.
It therefore cancels when two farthest-distance functions are compared along the same ray. In particular, for nonempty finite sets $B\subseteq A$,
\begin{equation}\label{eq:ray-loss-limit}
 F_A(\rho(t))-F_B(\rho(t)) \longrightarrow \log\frac{L_A(\xi)}{L_B(\xi)} \qquad(t\to\infty).
\end{equation}
The explicit error bounds justify this limit for every fixed ray and every fixed pair of finite sets.

\subsection{The exact loss identity}
The preceding calculation describes queries moving toward the ideal boundary. We now show that these queries capture the largest loss over all of $\HH^D$. The key observation is that a query can be moved away from one of its farthest input points without decreasing the coreset loss.

\begin{theorem}[Exact loss identity]\label{thm:boundary-loss}
For every pair of nonempty finite sets $B \subseteq A \subset \HH^D$, 
\begin{equation}\label{eq:exact-coreset-loss}
 \sup_{x\in\HH^D}\bigl(F_A(x)-F_B(x)\bigr) = \max_{\xi\in\SSph^{D-1}} \log\frac{L_A(\xi)}{L_B(\xi)}.
\end{equation}
\end{theorem}

\begin{proof}
The functions $L_A$ and $L_B$ are positive and continuous, and $L_B\le L_A$. Their logarithmic ratio therefore attains
a finite, nonnegative maximum on the compact sphere. Write 
\[
 \Delta = \max_{\xi\in\SSph^{D-1}} \log\frac{L_A(\xi)}{L_B(\xi)}.
\]

Fix $x \in \HH^D$. If $F_A(x)=0$, then $F_A(x)-F_B(x)=0\le\Delta$. Otherwise, choose $a\in A$ with $\Hd(x,a)=F_A(x)$. Extend the geodesic segment from $a$ through $x$ beyond $x$ to a unit-speed ray $\rho$ with $\rho(0)=x$, and let $\xi$ be its ideal endpoint. For every $t \ge 0$, $\Hd(\rho(t),a)=F_A(x)+t$. On the other hand, the triangle inequality gives $\Hd(\rho(t),a') \le t+\Hd(x,a') \le t+F_A(x)$ for every $a' \in A$. Thus $a$ remains farthest along the ray, and
\begin{equation}\label{eq:input-linear-growth}
 F_A(\rho(t))=F_A(x)+t.
\end{equation}
The same triangle inequality applied to $B$ gives
\begin{equation}\label{eq:coreset-at-most-linear-growth}
 F_B(\rho(t))\le F_B(x)+t.
\end{equation}
Subtracting yields $F_A(x)-F_B(x) \le F_A(\rho(t))-F_B(\rho(t))$. More generally, for $0\le s\le t$, $F_B(\rho(t))\le F_B(\rho(s))+t-s$. Together with~\eqref{eq:input-linear-growth}, this shows that the loss is nondecreasing along the entire ray.

Letting $t\to\infty$ and using~\eqref{eq:ray-loss-limit},
we obtain
\[
 F_A(x)-F_B(x) \le\log\frac{L_A(\xi)}{L_B(\xi)} \le\Delta.
\]
Taking the supremum over $x$ proves one inequality in~\eqref{eq:exact-coreset-loss}.

For the reverse inequality, choose a direction $\xi^\star$ attaining $\Delta$ and consider the standard ray $\gamma_{\xi^\star}$. Equation~\eqref{eq:ray-loss-limit} gives
\[
 F_A(\gamma_{\xi^\star}(t)) - F_B(\gamma_{\xi^\star}(t)) \longrightarrow\Delta.
\]
Every point on this ray at a finite parameter value is a valid query point. Hence the supremum of the finite-query losses is at least $\Delta$, proving the identity.
\end{proof}

\begin{remark}
The ideal-boundary maximum is attained, whereas the finite-query supremum need not be attained. A standard ray toward a maximizing ideal direction always approaches the supremum. Although the individual envelopes depend on the chosen hyperboloid coordinates, their maximum logarithmic ratio is intrinsic, by Theorem~\ref{thm:boundary-loss}.
\end{remark}

The following equivalent form is what we use in the coreset construction.

\begin{corollary}[Ideal-boundary characterization of additive coresets]\label{cor:additive-boundary-characterization}
Let $\varnothing\ne B\subseteq A\subset\HH^D$ be finite and let $\tau \ge 0$. Then
\[
 F_B(x)\ge F_A(x)-\tau \qquad\text{for every }x\in\HH^D
\]
if and only if
\begin{equation}\label{eq:boundary-score-test}
 L_B(\xi)\ge e^{-\tau}L_A(\xi) \qquad\text{for every }\xi\in\SSph^{D-1}.
\end{equation}
\end{corollary}

\begin{proof}
By Theorem~\ref{thm:boundary-loss}, the additive error is at most $\tau$ exactly when
\[
 \log\frac{L_A(\xi)}{L_B(\xi)}\le\tau \qquad\text{for every }\xi\in\SSph^{D-1}.
\]
Exponentiating gives~\eqref{eq:boundary-score-test}.
\end{proof}

Taking $A=P$ and $B=P_{\eps}$ reduces the coreset problem to approximating the ideal-boundary envelope uniformly. The next section normalizes these envelopes and uses a Euclidean $\delta$-kernel to obtain the required approximation.


\section{The Upper Bound}\label{sec:upper-bound}

By the Structure Theorem, it suffices to approximate the ideal-boundary envelope with small logarithmic loss in every direction. We do this by moving the minimum-enclosing-ball center to the origin, normalizing the spatial coordinates, and selecting a Euclidean $\delta$-kernel. The normalization controls both the diameter of the Euclidean point set and the minimum value of the envelope.

Let $P\subset\HH^D$ be finite and nonempty, and let $c$ and $R=\rad(P)$ be the center and radius of its minimum enclosing ball. Apply an isometry sending $c$ to $o$. This preserves all distances and both coreset guarantees. We work with the transformed input, denoting it again by $P$, and return the corresponding original input points at the end.

If $R=0$, the input consists of one point and is already an exact coreset. Henceforth assume $R>0$. Write
\[
 p=(\cosh r_p,\sinh r_p\,u_p), \qquad 0\le r_p\le R, \qquad u_p\in\SSph^{D-1}.
\]
For a point at $o$, the direction $u_p$ can be chosen arbitrarily. A point is \emph{active} if $r_p=R$.

\subsection{Balancing at the minimum enclosing ball}
The active directions cannot all point strictly toward the same side of the origin. Otherwise, moving the center slightly toward that side would decrease the enclosing radius. The following lemma makes this observation precise.

\begin{lemma}[Active-direction balance]\label{lem:active-balance}
Let $U_*=\{u_p:p\in P,\ r_p=R\}$. Then
\begin{equation}\label{eq:active-balance}
 \mathbf 0\in\conv U_*.
\end{equation}
Consequently, for every $\xi \in \SSph^{D-1}$, some active direction $u\in U_*$ satisfies $\langle \xi,u \rangle \le 0$.
\end{lemma}

\begin{proof}
Suppose $\mathbf 0\notin \conv U_*$. Let $y$ be the point of $\conv U_*$ closest to the origin. Such a point exists
because this convex hull is compact, and $y\ne\mathbf 0$. For every $u\in U_*$, $\langle y,u-y\rangle \ge 0$. Indeed, if this inner product were negative, moving a sufficiently small distance from $y$ toward $u$ would decrease the distance to the origin. Put
\[
 v=\frac{y}{\|y\|}, \qquad \mu=\|y\|>0.
\]
It follows that $\langle v,u\rangle \ge \mu$ for every $u \in U_*$.

Move the candidate center in direction $v$ by setting $c(t)=(\cosh t,\sinh tv)$ for $t>0$. For an active point $p=(\cosh R,\sinh Ru)$, the distance formula gives
\[
 \cosh\Hd(c(t),p)-\cosh R =(\cosh t-1)\cosh R -\sinh t\sinh R\,\langle v,u\rangle \le \sinh t\cosh R \left(\tanh\frac t2-\mu\tanh R\right).
\]
Here we used $\cosh t-1=\sinh t \tanh(t/2)$. Choose $t>0$ small enough that $\tanh(t/2)<\mu\tanh R$. Then every active point is at distance less than $R$ from $c(t)$.

If there are nonactive points, their finiteness gives a positive gap
\[
 \gamma=R-\max\{r_p:p\in P, r_p<R\}>0.
\]
Require also that $t<\gamma$. For every such point, $\Hd(c(t),p)\le t+r_p<R$. Thus all points of $P$ lie at distance strictly less than $R$ from $c(t)$, contradicting the minimum enclosing radius.

Finally, write the origin as a convex combination of active directions. Taking inner products with $\xi$ shows that their inner products with $\xi$ cannot all be positive.
\end{proof}

\subsection{A normalized Euclidean representation}
The spatial coordinate of a point at distance $r_p$ from $o$ has Euclidean norm $\sinh r_p$. To keep these coordinates
bounded even when $R$ is large, define 
\begin{equation}\label{eq:scaled-spatial-coordinates}
 \lambda=e^{-R}, \qquad z_p=e^{-R}\bar p=e^{-R}\sinh r_p u_p, \qquad Z=\{z_p:p\in P\}\subset\IR^D.
\end{equation}
This is an auxiliary Euclidean scaling used to select input points. Hyperbolic distances continue to be measured between the corresponding hyperboloid points.

For $\xi \in \SSph^{D-1}$, define
\begin{equation}\label{eq:g-function}
 g_\xi(z)=\sqrt{\lambda^2+\|z\|^2}-\langle\xi,z\rangle, \qquad z\in\IR^D.
\end{equation}
The hyperboloid equation gives
\[
 \sqrt{\lambda^2+\|z_p\|^2}=e^{-R}\cosh r_p, \qquad g_\xi(z_p)=e^{-R}\ell_p(\xi).
\]
Thus, writing $G_P(\xi)=\max_{p\in P}g_\xi(z_p)$, we have 
\begin{equation}\label{eq:envelope-scaled-g}
 G_P(\xi)=e^{-R}L_P(\xi).
\end{equation}

The first coordinate is determined by the spatial coordinates. After scaling, this relation becomes
\[
 e^{-R}p= \left(\sqrt{\lambda^2+\|z_p\|^2},z_p\right).
\]
Consequently, we can approximate the envelope by selecting points from the $D$-dimensional Euclidean set $Z$. The next lemma provides the two bounds needed for this approximation. The diameter bound controls the absolute error, and the positive lower bound on $G_P$ converts it into a logarithmic error.

\begin{lemma}[Normalization bounds]\label{lem:normalization-bounds}
Let $T=\min\{R,1\}$. Then
\begin{equation}\label{eq:G-bounds}
 G_P(\xi) \ge \frac{1}{2} \quad \text{for every } \quad \xi \in \SSph^{D-1},
\end{equation}
and
\begin{equation}\label{eq:Z-diameter}
 \diam(Z) \le 2T.
\end{equation}
\end{lemma}

\begin{proof}
Put
\[
 a=e^{-R}\sinh R=\frac{1-e^{-2R}}2, \qquad b=e^{-R}\cosh R=\frac{1+e^{-2R}}2.
\]
Fix $\xi \in \SSph^{D-1}$. By Lemma~\ref{lem:active-balance}, an active point has direction $u$ with $\langle\xi,u\rangle\le0$. Its scaled spatial coordinate is $au$, so
\[
 G_P(\xi)\ge g_\xi(au) = b-a\langle\xi,u\rangle \ge b\ge\frac12.
\]

Every $z_p$ has norm at most $a$, since $r_p\le R$. Therefore, $\diam(Z)\le2a=1-e^{-2R}$. Using $1-e^{-2R} \le 2R$ and $1-e^{-2R} \le 1$, we obtain $\diam(Z) \le 2\min\{R,1\}$.
\end{proof}

We now transfer a Euclidean convex-hull approximation to all the functions $g_\xi$ simultaneously.

\begin{lemma}[Convex transfer]\label{lem:convex-transfer}
Let $Z'\subseteq Z$ be nonempty, and put
\[
 G_{Z'}(\xi)=\max_{z\in Z'}g_\xi(z).
\]
If
\[
 d_{\mathrm H}^{\mathrm E}(\conv Z,\conv Z')\le\eta,
\]
then
\begin{equation}\label{eq:convex-transfer}
 0\le G_P(\xi)-G_{Z'}(\xi) \le 2\eta \qquad\text{for every }\xi \in \SSph^{D-1}.
\end{equation}
\end{lemma}

\begin{proof}
Write $f(z)=\sqrt{\lambda^2+\|z\|^2}=\|(\lambda,z)\|$, where the last expression is a Euclidean norm in $\IR^{D+1}$. The triangle inequality implies that $f$ is convex, and the reverse triangle inequality gives
\[
 |f(z)-f(w)| \le\|(\lambda,z)-(\lambda,w)\| =\|z-w\|.
\]
Since $z\mapsto-\langle\xi,z\rangle$ is linear and $\|\xi\|=1$, the function $g_\xi$ is convex and satisfies $|g_\xi(z)-g_\xi(w)|\le2\|z-w\|$. Thus $g_\xi$ is $2$-Lipschitz.

Choose $z^*\in Z$ maximizing $g_\xi$. By the Hausdorff-distance assumption, there is $y \in \conv Z'$ with $\|z^*-y\|\le\eta$. Write
\[
 y=\sum_i\alpha_i z_i', \qquad z_i'\in Z', \qquad \alpha_i \ge 0, \qquad \sum_i\alpha_i=1.
\]
The Lipschitz bound and convexity give
\[
 G_P(\xi)=g_\xi(z^*) \le g_\xi(y)+2\eta \le \sum_i\alpha_i g_\xi(z_i')+2\eta \le G_{Z'}(\xi)+2\eta.
\]
The inequality $G_{Z'}(\xi) \le G_P(\xi)$ follows from $Z'\subseteq Z$.
\end{proof}

Convexity is what allows us to select input points. The nearby point $y$ may lie inside $\conv Z'$ without belonging to $Z'$, but at least one selected point has a $g_\xi$-value as large as $g_\xi(y)$.

\subsection{The coreset construction}
We apply the Euclidean kernel theorem from the preliminaries, which provides a small set preserving all directional widths \cite{AgarwalHarPeledVaradarajan2004}.

\begin{theorem}[Upper bound]\label{thm:upper-bound}
Let $D \ge 2$ be fixed, let $P\subset\HH^D$ be finite and nonempty, and let $0<\eps\le1$. There exists a coreset
$P_{\eps}\subseteq P$ of size
\[
 |P_{\eps}|=O\left(\eps^{-(D-1)/2}\right)
\]
such that, for every $x \in \HH^D$,
\begin{equation}\label{eq:stronger-upper-error}
 0\le F_P(x)-F_{P_{\eps}}(x) \le \eps \min\{\rad(P),1\}.
\end{equation}
In particular, the same coreset satisfies
\[
 F_{P_{\eps}}(x)\ge F_P(x)-\eps \qquad\text{and}\qquad F_{P_{\eps}}(x)\ge(1-\eps)F_P(x)
\]
for every $x \in \HH^D$.
\end{theorem}

\begin{proof}
The case $R=\rad(P)=0$ is immediate. Assume $R>0$ and use the normalization above. Put $T=\min\{R,1\}$ and $\delta=\frac{\eps}{16}$. By Theorem~\ref{thm:euclidean-kernel}, the set $Z$ has a $\delta$-kernel $Z_{\eps}\subseteq Z$ with
\[
 |Z_{\eps}| = O\left(\delta^{-(D-1)/2}\right) = O\left(\eps^{-(D-1)/2}\right)
\]
and
\begin{equation}\label{eq:upper-kernel-hausdorff}
 d_{\mathrm H}^{\mathrm E}(\conv Z,\conv Z_{\eps}) \le \delta\,\diam(Z).
\end{equation}
Let $P_{\eps}$ contain the input points corresponding to $Z_{\eps}$. This correspondence is one-to-one, because the spatial coordinates determine a hyperboloid point uniquely and $e^{-R}>0$.

Use the same normalization, fixed by $P$, when evaluating the selected points. In particular, retain the same $R$ and $\lambda$, and define $G_{P_{\eps}}(\xi) = \max_{p\in P_{\eps}}g_\xi(z_p) = e^{-R}L_{P_{\eps}}(\xi)$. Lemmas~\ref{lem:normalization-bounds} and~\ref{lem:convex-transfer} give
\[
 0\le G_P(\xi)-G_{P_{\eps}}(\xi) \le 2\delta \diam(Z) \le 4\delta T.
\]
Since $G_P(\xi) \ge 1/2$ and $4\delta T \le 1/4$, we also have $G_{P_{\eps}}(\xi) \ge \frac{1}{4}$. We can therefore bound the logarithmic loss directly. Using $\log(1+u) \le u$ for $u \ge 0$, we obtain
\[
 \log\frac{L_P(\xi)}{L_{P_{\eps}}(\xi)} = \log\frac{G_P(\xi)}{G_{P_{\eps}}(\xi)} = \log\left( 1+\frac{G_P(\xi)-G_{P_{\eps}}(\xi)}{G_{P_{\eps}}(\xi)} \right) \le \frac{G_P(\xi)-G_{P_{\eps}}(\xi)}{G_{P_{\eps}}(\xi)} \le 16\delta T = \eps T.
\]
Equivalently,
\begin{equation}\label{eq:upper-envelope-approximation}
 L_{P_{\eps}}(\xi)\ge e^{-\eps T}L_P(\xi) \qquad\text{for every }\xi\in\SSph^{D-1}.
\end{equation}
The Structure Theorem~\ref{thm:boundary-loss} now gives
\[
 \sup_{x\in\HH^D} \bigl(F_P(x)-F_{P_{\eps}}(x)\bigr) \le \eps T.
\]
The loss is nonnegative because $P_{\eps}\subseteq P$, proving~\eqref{eq:stronger-upper-error}.

Since $T \le 1$, the additive loss is at most $\eps$. Moreover, the definition of the minimum enclosing radius
gives
\[
 F_P(x)\ge\min_{y\in\HH^D}F_P(y)=R\ge T.
\]
It follows that
\[
 F_{P_{\eps}}(x) \ge F_P(x)-\eps T \ge(1-\eps)F_P(x).
\]
Finally, return the original input points corresponding to the selected transformed points. The initial isometry preserves their cardinality and all distance guarantees.
\end{proof}

\begin{remark}\label{rem:no-scale-split}
The factor $\min\{R,1\}$ makes the construction work at every scale. For $R<1$, the diameter of $Z$ is $O(R)$, and the resulting distance loss is at most $\eps R$. For $R \ge 1$, the diameter of $Z$ is bounded by a constant, and the loss is at most $\eps$. The same normalization and kernel selection therefore provide both approximation guarantees for every input.
\end{remark}


\section{The Lower Bound}\label{sec:lower-bound}
We prove that the exponent in Theorem~\ref{thm:upper-bound} is optimal. The construction places input points on the hyperbolic sphere of radius one in well-separated directions. Each input point has an associated query at radius one in the opposite direction. The designated input point is at distance two from this query, whereas every other input point is closer by at least a constant times the square of the angular separation. An angular separation of $6\sqrt{\eps}$ makes every input point necessary for either approximation guarantee. A spherical packing provides $\Omega(\eps^{-(D-1)/2})$ such points.

\subsection{A spherical packing}
For $u,v\in\SSph^{D-1}$, define their angle by $\angle(u,v)=\arccos\langle u,v\rangle\in[0,\pi]$. A set $U\subseteq\SSph^{D-1}$ is \emph{$\theta$-separated} if $\angle(u,v)\ge\theta$ whenever $u,v\in U$ are distinct.

\begin{lemma}[Spherical packing]\label{lem:spherical-packing}
For every fixed $D\ge2$ and every $0<\theta\le1$, there is a finite $\theta$-separated set $U\subseteq\SSph^{D-1}$ with 
\begin{equation}\label{eq:packing-size}
 |U|=\Omega\left(\theta^{-(D-1)}\right).
\end{equation}
\end{lemma}

\begin{proof}
Choose a maximal $\theta$-separated set $U$. Such a set can be obtained by repeatedly adding an eligible point. This process terminates because compactness gives a finite cover of the sphere by sets of angular diameter less than $\theta$, each of which contains at most one selected point.

By maximality, the spherical caps of angular radius $\theta$ centered at the points of $U$ cover the sphere. Otherwise, a point outside all the caps could be added to $U$.

For $D=2$, each cap is an arc of length $2\theta$, so $2\pi\le2\theta|U|$ and the claim follows. For $D\ge3$, let $\sigma_k$ denote the surface area of the unit sphere $\SSph^k$. Spherical polar coordinates give the following bound on the area of each cap:
\[
 \sigma_{D-2}\int_0^\theta \sin^{D-2}\varphi d\varphi \le \sigma_{D-2}\int_0^\theta \varphi^{D-2}d\varphi = \frac{\sigma_{D-2}}{D-1}\theta^{D-1}.
\]
Here we used $\sin \varphi \le \varphi$. Since the caps cover $\SSph^{D-1}$,
\[
 \sigma_{D-1} \le |U| \frac{\sigma_{D-2}}{D-1}\theta^{D-1}.
\]
Rearranging proves~\eqref{eq:packing-size}.
\end{proof}

\subsection{A quadratic angular gap}
The next lemma bounds the distance loss when an input point is moved away from the direction opposite to a query. All points in the construction have hyperbolic radius one.

\begin{lemma}[Quadratic angular gap]\label{lem:angular-loss}
Let $u,v \in \SSph^{D-1}$, and put
\[
 p_v=(\cosh 1,\sinh 1\,v), \qquad x_u=(\cosh 1,-\sinh 1\,u).
\]
If $\alpha=\angle(u,v)$ and $d_{u,v}=\Hd(x_u,p_v)$, then
\begin{equation}\label{eq:angular-quadratic-loss}
 2-d_{u,v} \ge \frac{\tanh 1}{\pi^2}\alpha^2 \ge \frac{\alpha^2}{16}.
\end{equation}
\end{lemma}

\begin{proof}
The hyperboloid distance formula~\eqref{eq:hyperboloid-distance} gives
\[
 \cosh d_{u,v} = \cosh^2 1+\sinh^2 1\,\langle u,v\rangle = \cosh^2 1+\sinh^2 1\cos\alpha.
\]
Since $\cosh 2=\cosh^2 1+\sinh^2 1$, we have 
\begin{equation}\label{eq:cosh-angular-gap}
 \cosh 2-\cosh d_{u,v} = \sinh^2 1\,(1-\cos\alpha).
\end{equation}
Both points are at distance one from the origin, so the triangle inequality gives $d_{u,v}\le2$. Therefore,
\[
 \cosh 2-\cosh d_{u,v} = \int_{d_{u,v}}^2\sinh t dt  \le \sinh 2 (2-d_{u,v}).
\]
Combining this inequality with~\eqref{eq:cosh-angular-gap} and using $\sinh 2=2\sinh 1\cosh 1$, we obtain
\[
 2-d_{u,v} \ge \frac{\tanh 1}{2}(1-\cos\alpha).
\]
Concavity of sine on $[0,\pi/2]$ gives $\sin s \ge 2s/\pi$ on that interval. Since $0 \le \alpha \le \pi$, it follows that
\[
 1-\cos\alpha = 2\sin^2(\alpha/2) \ge \frac{2\alpha^2}{\pi^2}.
\]
This proves the first inequality in~\eqref{eq:angular-quadratic-loss}. Finally, $\tanh 1>3/4$ and $\pi^2<12$ imply
$\tanh 1/\pi^2>1/16$, proving the second inequality.
\end{proof}

In particular, angular separation at least $6\sqrt{\eps}$ gives a distance loss of at least $9\eps/4$. This exceeds both the additive tolerance $\eps$ and the multiplicative tolerance $2\eps$ at a query whose farthest distance is two.

\subsection{The hard instance}
\begin{theorem}[Lower bound]\label{thm:lower-bound}
For every fixed $D \ge 2$, there is a constant $c_D>0$ such that, for every $0<\eps \le 1/36$, there is a finite set $P^\star \subset \HH^D$ with
\[
 |P^\star| \ge c_D\eps^{-(D-1)/2}
\]
having the following property. If a nonempty set $P_{\eps}\subseteq P^\star$ satisfies either
\begin{equation}\label{eq:lower-additive-condition}
 F_{P_{\eps}}(x)\ge F_{P^\star}(x)-\eps \qquad\text{for every }x\in\HH^D
\end{equation}
or
\begin{equation}\label{eq:lower-multiplicative-condition}
 F_{P_{\eps}}(x)\ge(1-\eps)F_{P^\star}(x) \qquad\text{for every }x\in\HH^D,
\end{equation}
then $P_{\eps}=P^\star$.
\end{theorem}

\begin{proof}
Set $\theta=6\sqrt{\eps}$. The assumption $\eps \le 1/36$ ensures that $\theta \le 1$. By Lemma~\ref{lem:spherical-packing}, there is a $\theta$-separated set $U \subseteq \SSph^{D-1}$ with
\begin{equation}\label{eq:lower-U-size}
 |U| =\Omega\left(\theta^{-(D-1)}\right) =\Omega\left(\eps^{-(D-1)/2}\right).
\end{equation}
For each $u\in U$, define $p_u=(\cosh 1,\sinh 1u)$ and $x_u=(\cosh 1,-\sinh 1u)$, and let
\begin{equation}\label{eq:lower-construction}
 P^\star=\{p_u:u\in U\}.
\end{equation}
The points $p_u$ and $x_u$ lie on opposite rays from the origin, each at distance one from it. Hence
\begin{equation}\label{eq:lower-designated-distance}
 \Hd(x_u,p_u)=2.
\end{equation}
Every input point is at distance at most two from $x_u$ by the triangle inequality through the origin. Thus
\begin{equation}\label{eq:lower-farthest-value}
 F_{P^\star}(x_u)=2.
\end{equation}

For every $v \in U$ distinct from $u$, we have $\angle(u,v) \ge \theta$. Lemma~\ref{lem:angular-loss} therefore gives
\[
 \Hd(x_u,p_v) \le 2-\frac{\theta^2}{16} = 2-\frac{9}{4}\eps < 2-2\eps.
\]
In particular, $p_u$ is the unique farthest input point from $x_u$.

Suppose that $p_u\notin P_{\eps}$. Every selected point then has direction different from $u$, and hence
\begin{equation}\label{eq:lower-distance-gap}
 F_{P_{\eps}}(x_u) \le 2-\frac{9}{4}\eps < 2-2\eps.
\end{equation}
For the additive requirement, this implies $F_{P_{\eps}}(x_u) < 2-\eps = F_{P^\star}(x_u)-\eps$. For the multiplicative requirement, it implies $F_{P_{\eps}}(x_u) < 2-2\eps = (1-\eps)F_{P^\star}(x_u)$. Thus omitting $p_u$ violates each requirement at the same query point $x_u$.

The argument applies to every $u\in U$, so any coreset satisfying either requirement must contain every point of $P^\star$. Finally, distinct directions give distinct input points, and hence $|P^\star|=|U|$. Equation~\eqref{eq:lower-U-size} gives the claimed constant $c_D$.
\end{proof}

Together, Theorems~\ref{thm:upper-bound} and~\ref{thm:lower-bound} show that the optimal worst-case size is $\Theta\left(\eps^{-(D-1)/2}\right)$ for additive coresets, for multiplicative coresets, and for coresets satisfying both guarantees simultaneously.

\begin{remark}\label{rem:lower-scope}
The proof uses only queries on the hyperbolic sphere of radius one. The lower bound therefore holds even when queries are restricted to that sphere. As throughout the paper, the coreset consists of selected input points.

The construction also applies to two other error conventions. Define the hyperbolic diameter by
\[
 \diam_{\mathrm H}(P^\star) = \max_{p,q\in P^\star}\Hd(p,q).
\]
Since $\diam_{\mathrm H}(P^\star) \le 2$, omitting $p_u$ gives loss greater than $2\eps$, and therefore greater
than $\eps\diam_{\mathrm H}(P^\star)$. This proves the same lower bound for additive error $\eps \diam_{\mathrm H}(P^\star)$. For the alternative multiplicative requirement $F_{P^\star}(x) \le (1+\eps) F_{P_{\eps}}(x)$, Equation~\eqref{eq:lower-distance-gap} gives
\[
 (1+\eps)F_{P_{\eps}}(x_u) < (1+\eps)(2-2\eps) = 2(1-\eps^2) < 2 = F_{P^\star}(x_u).
\]
Thus that requirement also fails.
\end{remark}

\begin{remark}[Dimension one]\label{rem:D-one}
The parametrization $t\mapsto(\cosh t,\sinh t)$, for $t \in \IR$, identifies $\HH^1$ isometrically with the real line: the hyperboloid distance formula gives distance $|s-t|$ between parameters $s$ and $t$. Let $p^-$ and $p^+$ be the input points with the smallest and largest parameters. Then, for every $x \in \HH^1$,
\[
 F_P(x)=\max\{\Hd(x,p^-),\Hd(x,p^+)\}.
\]
These two points form an exact coreset; a singleton input needs only one point. The optimal worst-case size for $D=1$ is therefore $\Theta(1)$.
\end{remark}


\section{Algorithmic Construction}\label{sec:algorithm}
We describe the construction from Theorem~\ref{thm:upper-bound} and prove the running-time bound in Theorem~\ref{thm:algorithmic}. The algorithm computes the hyperbolic minimum enclosing ball, forms the normalized spatial point set, and selects a Euclidean $\delta$-kernel. Throughout this section, we represent a ball by its center $c$ and the value $\chi=\cosh R$, where $R$ is its radius. The algorithm does not need to evaluate $R$ itself.

\begin{algorithm}[htbp]
\caption{Hyperbolic farthest-point coreset}\label{alg:coreset}
\begin{algorithmic}[1]
\Require A nonempty finite set $P \subset \HH^D$ and $0<\eps \le 1$
\Ensure A coreset $P_{\eps} \subseteq P$ satisfying
         \eqref{eq:stronger-upper-error}
\State $(c,\chi) \gets \Call{HyperbolicMEB}{P}$
\If{$\chi=1$}
    \State \Return $\{p\}$ for any $p \in P$
\EndIf
\State $\lambda\gets 1/(\chi+\sqrt{\chi^2-1})$
\State Compute an isometry $\varphi_c$ sending $c$ to $o$
\State $Z\gets\varnothing$
\For{each $p\in P$}
    \State $(\widetilde p_0,\overline{\widetilde p})
           \gets\varphi_c(p)$
    \State $z_p\gets\lambda\,\overline{\widetilde p}$
    \State Insert $z_p$ into $Z$ and store
           $\operatorname{orig}(z_p)\gets p$
\EndFor
\State $\delta\gets\eps/16$
\State $Z_{\eps}\gets\Call{EuclideanKernel}{Z,\delta}$
       \Comment{Theorem~\ref{thm:euclidean-kernel}}
\State \Return
       $P_{\eps}=\{\operatorname{orig}(z):z\in Z_{\eps}\}$
\end{algorithmic}
\end{algorithm}

Here $\operatorname{orig}(z)$ stores the original input point represented by $z$. This correspondence is one-to-one because the isometry is bijective, $\lambda>0$, and a hyperboloid point is uniquely determined by its spatial coordinate. The normalization is exactly the one used in the upper-bound proof. Consequently, Theorem~\ref{thm:upper-bound} proves correctness and the output-size bound
\[
 |P_{\eps}|=O\left(\eps^{-(D-1)/2}\right).
\]
We next justify the minimum-enclosing-ball computation and the remaining algebraic operations.

\subsection{The minimum enclosing ball as an LP-type problem}
For a nonempty finite set $S \subset \HH^D$, define
\[
 \omega(S)=\cosh\bigl(\rad(S)\bigr), \qquad \omega(\varnothing)=0.
\]
Thus every nonempty constraint set has value at least one. The constraints are the input points. A point $p$ violates
$S$ when $\omega(S\cup\{p\})>\omega(S)$. A \emph{basis} is a constraint set for which every proper subset has a smaller objective value. The \emph{combinatorial dimension} is the maximum possible basis size.

We verify the two LP-type properties, monotonicity and locality, and bound this dimension. The algorithmic framework is due to Sharir and Welzl~\cite{SharirWelzl1992}; Welzl~\cite{Welzl1991} gives the randomized incremental approach for smallest enclosing balls.

\begin{lemma}\label{lem:meb-lp-type}
The hyperbolic minimum-enclosing-ball problem is an LP-type problem of combinatorial dimension at most $D+1$. With constant-time violation tests and constant-time basis computations on a bounded number of constraints, its solution can be found in randomized expected $O(n)$ time for fixed $D$.
\end{lemma}

\begin{proof}
If $S\subseteq T$, every ball containing $T$ contains $S$, so $\omega(S)\le\omega(T)$. This is monotonicity.

For locality, first let $S$ be nonempty and write $\mathcal B(S)$ for its unique minimum enclosing ball. A point $p$ violates $S$ precisely when $p\notin\mathcal B(S)$. One implication is immediate. For the other, if the objective did not increase after adding a point outside $\mathcal B(S)$, the new optimal ball would also be an optimal ball for $S$. Uniqueness would force it to equal $\mathcal B(S)$, a contradiction.

Now suppose $S\subseteq T$ and $\omega(S)=\omega(T)$. If they are nonempty, $\mathcal B(T)$ is also an optimal ball for $S$, so uniqueness gives $\mathcal B(S)=\mathcal B(T)$. Hence any new point violates $S$ if and only if it violates $T$. If $S=\varnothing$, equality of the objective values forces $T=\varnothing$, and the same conclusion is immediate. This proves locality.

To bound the basis size, let $\mathcal C$ be a nonempty basis and put $R=\rad(\mathcal C)$. If $R=0$, one point has the same objective value, so $|\mathcal C|=1$. Suppose $R>0$, and move the optimal center to $o$. Lemma~\ref{lem:active-balance} gives $0\in\conv U_*$ for the active directions. Carath\'eodory's theorem supplies at most $D+1$ active directions $u_1,\ldots,u_m$ and coefficients $\alpha_i\ge0$ such that
\[
 \sum_{i=1}^m\alpha_i=1, \qquad \sum_{i=1}^m\alpha_i u_i=0.
\]
Let $p_i=(\cosh R,\sinh R\,u_i)$ be the corresponding points. For any candidate center $x=(\cosh t,\sinh t\,v)$, the distance formula gives
\[
 \sum_{i=1}^m\alpha_i\cosh\Hd(x,p_i) = \cosh t\cosh R -\sinh t\sinh R \left\langle v,\sum_{i=1}^m\alpha_i u_i\right\rangle = \cosh t\cosh R \ge \cosh R.
\]
Thus at least one selected point has distance at least $R$ from $x$. The selected points have minimum enclosing radius exactly $R$, since the ball centered at $o$ also encloses them. By minimality of the basis, $\mathcal C$ consists of these points. Therefore $|\mathcal C|\le D+1$; the empty basis also satisfies this bound.

The standard randomized LP-type algorithm now gives expected linear time for fixed combinatorial dimension under the stated primitive-operation assumptions~\cite{SharirWelzl1992}.
\end{proof}

\subsection{Algebraic primitives}
We use the real-RAM model with unit-cost arithmetic, comparisons, and square roots. In fixed dimension, Lorentz inner products and fixed-size linear systems also take constant time. We show that these operations suffice for the primitives in
Lemma~\ref{lem:meb-lp-type}.

A violation test for a nonempty constraint set is simply $\langle c,p\rangle_{\mathrm L}>\chi$, where $(c,\chi)$ represents its minimum enclosing ball. Every point violates the empty constraint set.

For completeness, the following lemma gives a constant-size solver that also handles degenerate configurations. A basis update may involve a basis together with one additional point, so it is useful to allow up to $D+2$ constraints.

\begin{lemma}\label{lem:small-meb}
For fixed $D$, the minimum enclosing ball of a nonempty set of at most $D+2$ hyperboloid points can be computed in constant time using arithmetic, comparisons, and square roots.
\end{lemma}

\begin{proof}
Let $S$ be the nonempty set to be enclosed. Enumerate every nonempty subset $I=\{p_1,\ldots,p_m\}\subseteq S$ with $m\le D+1$. Its Lorentz Gram matrix form is $M_{ij}=\langle p_i,p_j\rangle_{\mathrm L}$. If $M$ is singular, skip this subset. Otherwise, solve $M\beta=\mathbf 1$, where $\mathbf 1$ is the vector of $m$ ones. Keep only solutions with $\beta_i>0$ for every $i$.
Set
\[
 h=\sum_{i=1}^m\beta_i p_i, \qquad \sigma=\sum_{i=1}^m\beta_i, \qquad c_I=\frac{h}{\sqrt{\sigma}}, \qquad \chi_I=\frac{1}{\sqrt{\sigma}}.
\]
We have
\[
 \langle h,h\rangle_{\mathrm L} = \beta^{\mathsf T}M\beta = \sigma>0.
\]
Also $h_0>0$, because all $\beta_i$ and all first coordinates $(p_i)_0$ are positive. Hence $c_I \in \HH^D$, and
$\langle c_I,p_i\rangle_{\mathrm L}=\chi_I$ for every $i$. Retain the candidate if $\langle c_I,p\rangle_{\mathrm L}\le\chi_I$ for every $p \in S$. Return a retained candidate with minimum $\chi_I$.

We verify that the true optimal ball occurs among these candidates. Let its center and value be $c^\star$ and $\chi^\star$. The radius-zero case is recovered by any singleton. Otherwise, move $c^\star$ to the origin and choose
an inclusion-minimal set of active directions whose convex hull contains zero. These directions are affinely independent and have strictly positive balancing coefficients $\alpha_i$. Indeed, an affine dependence would let us vary the coefficients while preserving their sum and weighted sum, until one coefficient became zero, contradicting minimality.

The corresponding hyperboloid points are linearly independent. In the original coordinates, their balance condition is $\sum_i\alpha_i p_i=\chi^\star c^\star$. Their span therefore contains $c^\star$. The Lorentz form restricted to this span is nondegenerate: after sending $c^\star$ to $o$, the span is the direct sum of the line through $o$ and a spatial subspace, on which the form is negative definite. Thus their Gram matrix is nonsingular. The vector $\beta_i = \frac{\alpha_i}{(\chi^\star)^2}$ solves $M\beta=\mathbf 1$ and has positive entries. The construction consequently recovers $c_I=c^\star$ and $\chi_I=\chi^\star$.

Every retained candidate encloses $S$, so choosing the smallest value gives the optimum. There are only constantly many subsets in fixed dimension, and each calculation has constant size. An inclusion-minimal defining subset can likewise be found by testing the subsets of $S$.
\end{proof}

The isometry used by Algorithm~\ref{alg:coreset} can also be evaluated algebraically. For $c=(c_0,\bar c)$, define
\begin{equation}\label{eq:algorithm-isometry}
 \varphi_c(p_0,\bar p) = \left( c_0p_0-\langle\bar c,\bar p\rangle, \bar p+ \left( \frac{\langle\bar c,\bar p\rangle}{1+c_0}-p_0 \right)\bar c \right).
\end{equation}
Substitution shows that this linear map preserves the Lorentz form and sends $c$ to $o$. For $p\in\HH^D$, its first coordinate is $\langle c,p\rangle_{\mathrm L}=\cosh\Hd(c,p)>0$, so it preserves the upper sheet and is a hyperbolic isometry. Computing its coefficients and applying it to one point take constant time for fixed $D$.

Finally, since $\chi=\cosh R$ and $R\ge0$, $\chi+\sqrt{\chi^2-1} = \cosh R+\sinh R=e^R$. Thus the algorithm computes the required normalization factor as $\lambda=\frac{1}{\chi+\sqrt{\chi^2-1}}=e^{-R}$ using only the allowed operations.

\subsection{Running time}
\begin{proof}[Proof of Theorem~\ref{thm:algorithmic}]
Let $|P|=n$, and fix $D\ge3$ and $\alpha>0$. Lemmas~\ref{lem:meb-lp-type} and~\ref{lem:small-meb} give randomized expected $O(n)$ time to compute $(c,\chi)$. If $\chi=1$, the algorithm returns immediately. Otherwise, applying~\eqref{eq:algorithm-isometry}, forming all normalized spatial coordinates, and storing their input pointers take $O(n)$ time.

By Theorem~\ref{thm:euclidean-kernel}, the deterministic Euclidean kernel algorithm of Arya, da Fonseca, and Mount~\cite[Theorem~1.1]{AryaDaFonsecaMount2017}, with $\delta=\eps/16$, takes time
\[
 O\left( n\log\frac{2}{\delta} + \delta^{-(D-1)/2-\alpha} \right) = O\left( n\log\frac{32}{\eps} + \eps^{-(D-1)/2-\alpha} \right) = O\left( n\log\frac{2}{\eps} + \eps^{-(D-1)/2-\alpha} \right).
\]
Here $D$ and $\alpha$ are fixed. Returning the stored input points takes time linear in the output size and is absorbed by this bound. This proves the claimed expected running time.

The exponent slack $\alpha$ comes entirely from the Euclidean kernel algorithm and does not change the coreset size. Every step after the minimum enclosing ball computation is deterministic, so the expectation arises only from the randomized LP-type computation.
\end{proof}

\begin{remark}[Computational model]\label{rem:machine-model}
The algorithm stores $(c,\chi)$ rather than $(c,R)$. The formulas above show that it does not need to evaluate $\exp$, $\log$, or $\arcosh$. The real-RAM assumptions are used only for the construction-time bound; the coreset size and approximation guarantees are geometric statements.
\end{remark}


\section{Conclusion}\label{sec:conclusion}
We have shown that, for every fixed dimension $D \ge 2$, the optimal worst-case size of a coreset for hyperbolic
farthest-point queries is $\Theta\left(\eps^{-(D-1)/2}\right)$. A single coreset $P_{\eps}\subseteq P$ satisfies
\[
 F_{P_{\eps}}(x)\ge F_P(x)-\eps \qquad\text{and}\qquad F_{P_{\eps}}(x)\ge(1-\eps)F_P(x)
\]
for every query point $x \in \HH^D$. The lower bound applies even when either approximation guarantee is required separately.

The main geometric ingredient is the exact ideal-boundary envelope identity. It expresses the uniform farthest-distance loss over all finite query points as the maximum logarithmic loss between the corresponding envelopes on the ideal boundary. Minimum-enclosing-ball normalization then allows Euclidean convex-hull approximation to control these envelope losses. The spatial graph representation of the hyperboloid keeps the Euclidean kernel construction in $\IR^D$, yielding the upper bound. A spherical packing with angular separation proportional to $\sqrt{\eps}$, together with antipodal queries, gives the lower bound.

For every fixed $D \ge 3$ and every fixed $\alpha>0$, the construction can be implemented in the real-RAM model in expected time $O\left( n\log\frac{2}{\eps} + \eps^{-(D-1)/2-\alpha} \right)$ with $n=|P|$. The expectation arises only from the randomized minimum-enclosing-ball computation, while the exponent slack $\alpha$ comes from the deterministic Euclidean kernel algorithm.

A natural direction is to study farthest-point coresets in Siegel spaces, whose upper half-space and disk descriptions use complex symmetric matrices. Nielsen~\cite{Nielsen2020SiegelKlein} studied the Hilbert geometry of the Siegel disk and algorithms for approximating minimum enclosing balls in these matrix domains. Extending coreset results to such settings requires specifying both the metric and the approximation guarantee. Under the standard invariant Riemannian metric,
Siegel spaces of matrix size at least two contain isometrically embedded Euclidean planes~\cite[Section~3]{LopezEtAl2021}. Scaling regular polygons in such a plane makes arbitrarily many input points indispensable for
a fixed absolute additive error $\eps$, ruling out a bound that depends only on the dimension and $\eps$. This motivates multiplicative guarantees, additive error relative to the input diameter, or bounds with explicit dependence on the input radius. The Hilbert metric on the Siegel--Klein disk requires a separate analysis. In both settings, it would be useful to determine how explicit boundary representations can support efficient coreset constructions.

Further questions concern streaming and dynamic maintenance of coresets of comparable size. Our construction depends on a minimum enclosing ball that may change after insertions or deletions, so an efficient maintenance algorithm must account for changes in the normalization. It would also be interesting to understand whether more general summaries, including representatives outside the input, permit smaller representations under a specified query and error model.

\bibliographystyle{plain}
\bibliography{references}

\end{document}